# Monitoring the refractive index of tissue models using light scattering spectroscopy

Michelle Bailey,[1] Benjamin Gardner,[1] Martina Alunni-Cardinali,[2] Silvia Caponi,[3] Daniele Fioretto,[2] Nick Stone,[1] and Francesca Palombo[1,*]

[1]School of Physics and Astronomy, University of Exeter, Exeter, EX4 4QL, UK
[2]Department of Physics and Geology, University of Perugia, Perugia, Italy
[3]CNR-IOM - Istituto Officina dei Materiali - Research Unit in Perugia, c/o Dept. of Physics and Geology, University of Perugia, Perugia, Italy

*Corresponding author: f.palombo@exeter.ac.uk

## Abstract

In this work, we report the application of Raman microspectroscopy for analysis of the refractive index of a range of tissue phantoms. Using both a custom-developed setup with visible laser source and a commercial micro-spectrometer with near infrared laser, we measured the Raman spectra of gelatin hydrogels at various concentrations. By building a calibration curve from measured refractometry data and Raman scattering intensity for different vibrational modes of the hydrogel, we determined the refractive indices of the gels from their Raman spectra, with approximately the same accuracy as that of refractometry measurements with an Abbe refractometer. This work highlights the importance of a correlative approach through Brillouin-Raman microspectroscopy for the mechano-chemical analysis of biologically relevant samples.

## Introduction

Gelatin hydrogels derived from denatured collagen [1] constitute a simple model to investigate the physical properties of connective tissue. Gelatin is characterised in a large part by the presence of water, the medium with low compressibility in all biological processes, interspersed within a network of protein molecules, conferring the shear load-bearing property to the system. As such, it is a stable, low-cost, safe and easy-to-prepare system for optical and biomechanical testing.
Brillouin spectroscopy is a vibrational spectroscopy technique with a unique potential for mechanobiology and biomedical sciences [2-3]. It is based on the inelastic light scattering effect where incident light is scattered by thermally driven acoustic waves, or 'phonons', which propagate as material density fluctuations resulting in periodic changes in refractive index [4]. Information on biomechanics is provided both by measurement of the frequency shift, which gives access to the longitudinal elastic modulus, and the linewidth of the Brillouin peak, which yields the attenuation of the acoustic wave and is a measure of the apparent viscosity. Brillouin microspectroscopy (BM) has proved to be an effective probe of biomechanics (more specifically, micro-viscoelasticity) in a range of biological samples, including live cells [5-6] and organisms [7-8], human tissue sections [9-10] and cornea [11]. Despite the clear advantages of BM as a non-destructive, contactless probe of micro-biomechanics, it is truly the correlative approach with complementary techniques alongside BM that is most beneficial in enhancing the specificity of the measurements as it facilitates access to the full information contained within Brillouin spectra. Raman spectroscopy is a promising correlative technique. It provides valuable information on the chemical composition and structure of materials through the inelastic scattering of light from molecular vibrations; hence, it is label-free and chemically specific. We first proposed to interface Brillouin and Raman microspectroscopy [12] and then realised the first high-contrast Brillouin-Raman microscope [13], which enables simultaneous measurement of the micro-mechanical and chemical properties of samples [14]. Brillouin and Raman spectroscopy are 'sister' techniques, sharing a common optical arrangement and similar light scattering effects, occurring at adjacent frequency scales. In Raman spectroscopy, as well as BM, the signal intensity is linearly proportional to the concentration or density of the scattering species [15-16]. A less explored dependence is that between signal intensity and refractive index, which we aim to investigate in this work.

In the emerging BioBrillouin community, various efforts have been made to assess the refractive indices of samples measured by confocal scanning BM, in order to decouple the optical from the mechanical effects, which contribute to the overall Brillouin line shape. These have been reviewed in recent works [1-2, 17], so we only recall here that traditional refractometry is insufficient for measuring the refractive index of samples with the micrometric resolution that is required in BM studies. Recent advances in phase imaging have enabled quantitative phase imaging [7], holographic phase microscopy [18] and optical diffraction tomography (ODT) [19-20] to be implemented alongside BM. Brillouin spectroscopy itself has also been used to determine the refractive index of samples by utilising two different scattering geometries or angles [21-23], however routine use of these approaches has not yet come to fruition.

In this work, Raman microspectroscopy was successfully applied to gelatin hydrogels, used as biological tissue models, to monitor the refractive index of the gels with chemical specificity. A calibration model based on Raman band intensities provides access to the refractive index to approximately the same accuracy as measurements conducted with an Abbe refractometer [1], a finding that can open the full potential of Brillouin imaging in biomedical and life sciences.

**Experimental**

Type B gelatin (denatured collagen) was prepared to concentrations between 4 and 18 % w/w as previously described [1, 24-25]. The Brillouin spectrum and refractive index of all gelatin samples were measured by high-contrast Brillouin microscopy and Abbe refractometry, with 532 nm and D line (589 nm) light sources, respectively, as part of a previous work [1, 24].

The Raman spectra of gelatin at different concentrations were collected using two different systems. A Renishaw inVia confocal microscope with long working distance 50× (NA 0.50) objective and using an 830 nm laser, was employed for measurements in the 'fingerprint' region. Each sample, prepared in a cylindrical mould, was transferred onto a Raman-grade calcium fluoride substrate (Crystran, UK) and analysed by Raman microspectroscopy. The backscattered light from the sample was dispersed through a 600 lines/mm grating onto a Renishaw deep depletion CCD camera. Raman spectra were acquired with an exposure time of 7 s per spectrum and 32 accumulations, in the range 372–2345 cm$^{-1}$. Three spectra were collected at different locations within the sample for all gel concentrations. WiRE v. 4.0 software was used for data acquisition.

A microscope system equipped with a 20× (NA 0.42) objective and using a 532 nm laser and a Horiba iHR320 Triax Raman spectrometer was used for measurements in the high wavenumber region. The backscattered light from the sample held in a sealed glass cuvette was dispersed through a 600 lines/mm grating onto a CCD camera. Raman spectra of the gels were measured with an exposure time of 1 s and 60 accumulations, in the range 171–3843 cm$^{-1}$. Five spectra were collected at different locations within the sample for all gel concentrations. LabSpec5 software was used for data acquisition.

Spectra were processed in MATLAB using custom written scripts. Firstly, spectral pre-processing (Fig. 1) was performed in three steps: (1) cosmic ray removal, (2) baseline subtraction using an asymmetric least squares method [26], and (3) normalisation of each spectrum to its Euclidian norm.

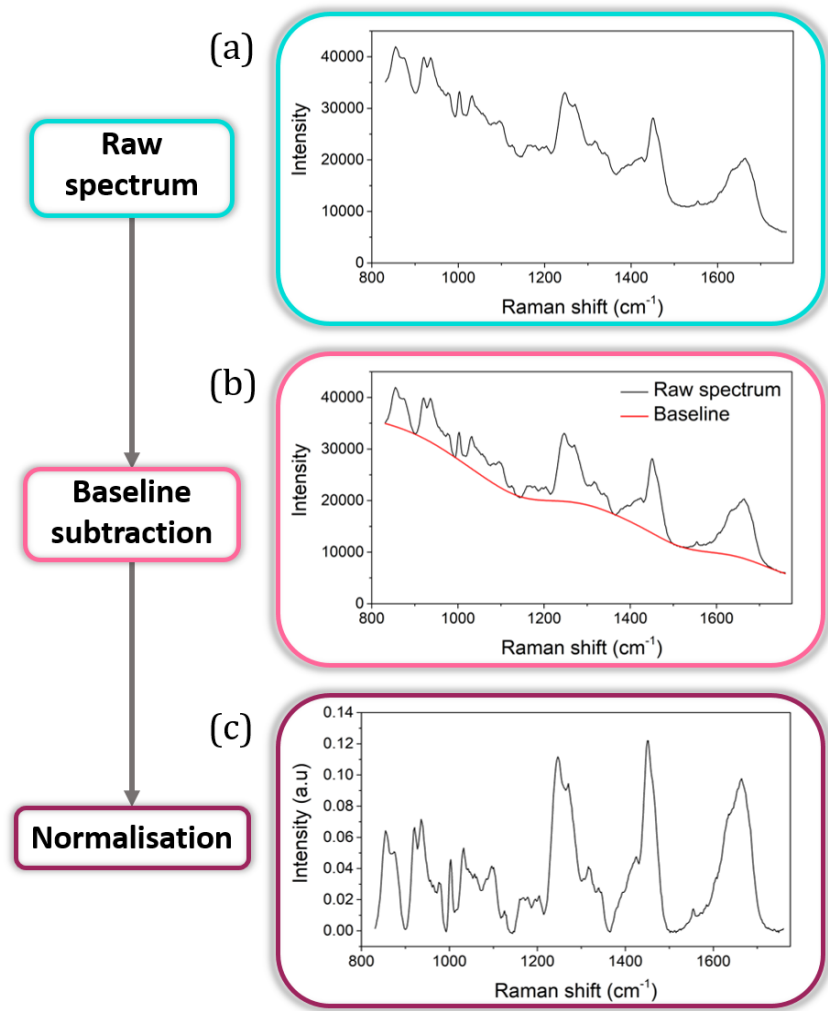

**Figure 1.** Data processing for the Raman spectrum of an 18% gelatin hydrogel. (a) Raw spectrum. (b) Baseline (red line) is determined by asymmetric least squares fitting to the raw spectrum (black line). (c) Spectrum is normalised through division by its Euclidian norm.

## Results and Discussion

Figure 2 shows the evolution in the spectra of the gels as a function of concentration across the 'fingerprint' and CH stretching regions.

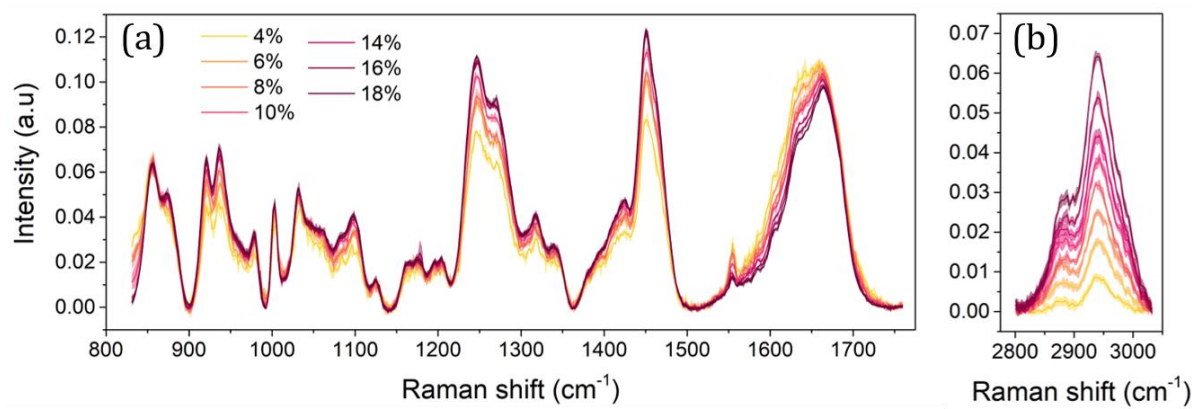

**Figure 2.** Normalised Raman spectra of gelatins measured across (a) the 'fingerprint' and (b) C–H stretching region. Each spectrum is an average of (a) three or (b) five measurements, pre-processed and analysed as described in the text.

It can be seen that there is a clear trend in the change of scattering intensity such that, for example, the CH stretching peaks increase with increasing concentration (Fig. 2b).

The gelatin spectra in the 'fingerprint' region were mean-centred and analysed by Principal Component Analysis (PCA) to determine the spectral regions responsible for the variance of the dataset. In PCA, the principal components are ranked in such a way that the first component accounts for the highest percentage of the total variance. The first principal component (PC1) accounted for 89% of the total data variance and the loadings (Fig. 3a) highlight the spectral regions where this variation is observed. The corresponding score plot simply represents the trend of the variation described by the loadings versus concentration (Fig. 3b).

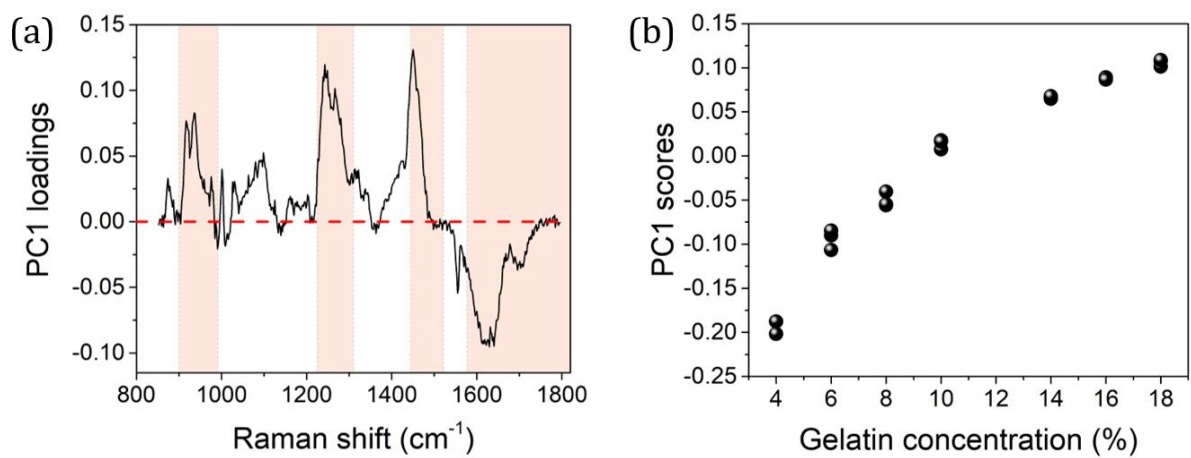

**Figure 3.** PCA applied to the Raman spectra of gelatins in the 'fingerprint' region. (a) First principal component (PC1) loading and (b) score plot. Shading in (a) denotes the spectral regions that express most variance. Tight clustering between repeated measurements at each concentration is observed in (b).

The positive loadings (Fig. 3a) result from those peaks that increase in intensity with increasing gel concentration, while the negative loadings are for those signals that decrease in intensity with increasing concentration. The former signals are assigned to the protein component of the gels, while the latter is mainly ascribed to water (see below). It follows that gels with higher water content (4–8 % w/w) have negative PC1 scores (Fig. 3b), while those with lower water content (≥10 %) have positive scores. Fig. 3a highlights the spectral regions where signals contribute most variance in the dataset. Among these we can identify the range 898–988 cm$^{-1}$, which contains a doublet at 922 and 938 cm$^{-1}$ (C–C stretching of the proline ring and plausibly C–C stretching of the protein backbone [27-28]) and a small peak at 980 cm$^{-1}$ (arginine [29]). The signals in this region are sensitive to the presence of 'bound' water within the hydrogel [30]; they are indeed found to increase with increasing gel concentration as the number of binding sites increases. An increase in bound water with increasing concentration has already been derived in our previous Brillouin study of gelatin hydrogels [1]. The range 1216–1300 cm$^{-1}$ presents a doublet at 1248 and 1271 cm$^{-1}$ (amide III [27]), whilst the range 1431–1507 cm$^{-1}$ corresponds to $CH_3$ and $CH_2$ deformations [27, 31]. The range between 1562 and 1800 cm$^{-1}$ presents contributions from both protein (amide I [27, 31] centred at 1665 cm$^{-1}$ assigned to disordered protein structure [32-33], with a shoulder at 1635 cm$^{-1}$ associated with denatured triple helices [33]) and water (bending mode at 1635 cm$^{-1}$ [34-35]). Highly hydrated gelatin is expected to be more disordered than gels of lower water content, where a larger proportion of alpha helices are present [33]. In addition to the 'fingerprint' range, the C–H stretching band [36] which present two peaks at 2885 and 2940 cm$^{-1}$ (symmetric and antisymmetric $CH_2$ stretches, respectively) was used in the analysis.
Raman signals were integrated with respect to frequency shift, and the intensities obtained were used to build the calibration plots for refractive index analysis. The bands analysed in this way were the main protein resonances in the 'fingerprint' and high wavenumber regions, i.e. amide I at 1665 cm$^{-1}$ and C–H stretching at 2800–

3040 cm$^{-1}$. These band intensities display a linear dependence on gel concentration across the entire range studied here. Similarly, the refractive index of the gels presents a linear dependence on concentration (see ref. [1]). This enables a model to be constructed, where the integrated intensities of amide I and CH bands are plotted versus refractive index (Fig. 4a and Fig. 5a).

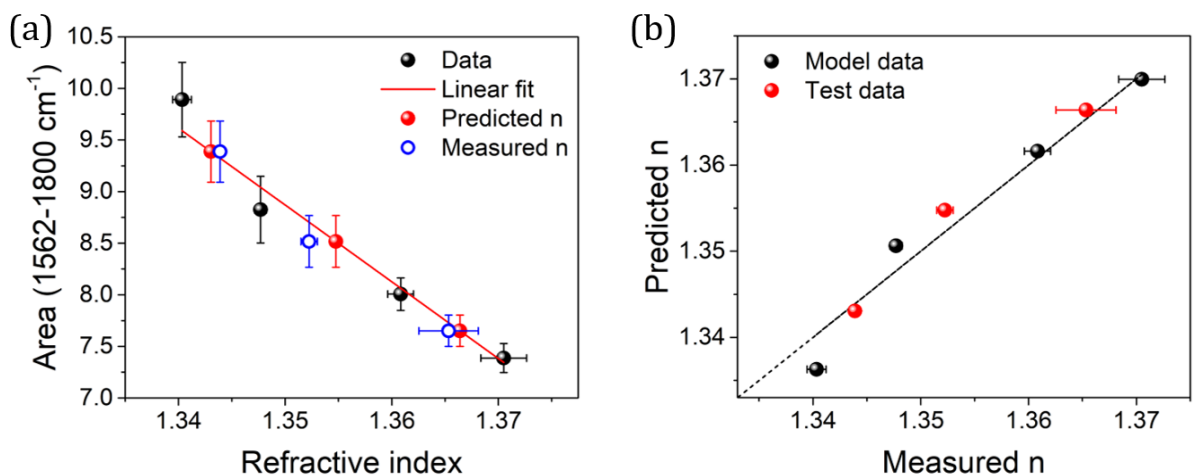

**Figure 4.** (a) Plot of amide I integrated intensity versus refractive index. Red line denotes a linear fit of the dataset used as model data (black dots): R$^2$=0.96. (b) Predicted vs measured refractive indices. Black dots denote data from which the model was calculated and red dots correspond to test data, where refractive index was determined from the model. Dashed line ($y = x$) serves as a guide for the eye. Error bars denote the standard deviation.

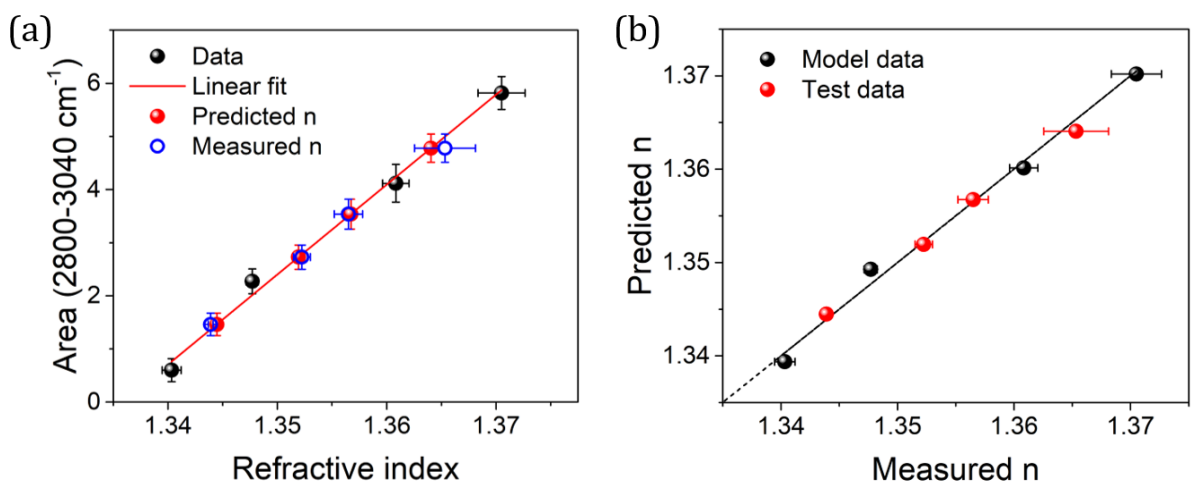

**Figure 5.** (a) Plot of CH stretching integrated intensity versus refractive index. Red line denotes a linear fit of the dataset used as model data (black dots): R$^2$=0.99. (b) Predicted vs measured refractive indices. Black dots denote data from which the model was calculated and red dots correspond to test data, where refractive index was determined from the model. Dashed line ($y = x$) serves as a guide for the eye. Error bars denote the standard deviation.

In the model, the data are split into two subsets: half of the samples were used as model data to derive the fit and the other half were used as test samples to determine the refractive index. The data points selected as model data corresponded to 4 %, 8 %, 14 % and 18 % gels (black filled circles in Fig. 4a and Fig. 5a). A linear fit was applied to these data (red line), and calibration functions were derived for the amide I: $y = 109.3137 - 74.402n$, and the CH stretch: $y = 169.351n - 226.226$, where $y$ represents the integrated intensity of the peak and $n$ the corresponding refractive index of the gel. The refractive indices derived from the model (red filled circles) using the measured integrated intensity on the x axis were then compared with those obtained with an Abbe refractometer [1] (blue empty circles). Fig. 4b and Fig. 5b show that there is a close correspondence between predicted and measured values of $n$, confirmed by RMSE (root mean square error) values of 0.002 and 0.0009 for the predicted vs measured refractive indices determined by the amide I and CH stretching modes, respectively. This indicates that the model is capable of predicting the refractive index of the gelatin hydrogels with high accuracy. Table I lists all results from this analysis.

**Table I.** Refractive indices derived from Abbe refractometry [1] and Raman measurements using the calibration model.

| Gel concentration (%) | Measured $n$ (± SD)[a] | Predicted $n$ (± difference)[b] | |
|---|---|---|---|
| | | Amide I | ν(CH) |
| 4 | 1.3403 (± 0.0009) | 1.3363 (± 0.004) | 1.3394 (± 0.0009) |
| 6 | 1.3403 (± 0.0002) | 1.3431 (± 0.0008) | 1.3445 (± 0.0006) |
| 8 | 1.3477 (± 0.0006) | 1.3506 (± 0.003) | 1.3493 (± 0.0016) |
| 10 | 1.3523 (± 0.0008) | 1.3548 (± 0.003) | 1.3519 (± 0.0003) |
| 12 | 1.356 (± 0.001) | - | 1.3567 (± 0.0002) |
| 14 | 1.361 (± 0.001) | 1.3616 (± 0.0008) | 1.3602 (± 0.0007) |
| 16 | 1.365 (± 0.003) | 1.3664 (± 0.0011) | 1.3641 (± 0.0013) |
| 18 | 1.370 (± 0.002) | 1.3699 (± 0.0006) | 1.3702 (± 0.0003) |

Shading denotes data which were used for calibration.
[a] Standard deviation derived from five measurements at each concentration.
[b] Difference between refractive indices measured by Abbe refractometry and those determined from Raman spectroscopy.

A very good estimation of the refractive index is found using this method, with predicted values of $n$ being within 0.02-0.3 % of the measured values. Differences between measured and predicted values were of the same order of magnitude as the standard deviation of the measurements performed with an Abbe refractometer

(Table I). Prediction based on CH stretching analysis was generally more accurate than that based on amide I, as can be expected because the CH stretching modes are exclusively protein modes, whilst the amide I band contains a contribution from the water bending mode that has an opposite trend with increasing concentration.

**Conclusion**

In summary, we have demonstrated that Raman spectroscopy can be applied to assess the refractive index of biologically relevant samples through appropriate calibration based on integrated peak intensity. The refractive index of gelatin hydrogels displays a linear dependence with concentration and a similar linear relation is observed for the integrated intensity of the amide I (1562–1800 cm$^{-1}$) and CH stretching region (2800–3040 cm$^{-1}$). Using this relation, we have shown that the refractive index can be predicted to approximately the same accuracy as that of Abbe refractometry measurements. This is an important result that further substantiates implementations where Raman spectroscopy is applied alongside Brillouin microscopy, as it provides complementary information on the chemical and structural properties of the sample as well as indirectly its refractive index.
There are limitations of this work to note. In fact, the refractive index assessment was performed on a simple model of a biological sample, whereas real specimens such as human tissues are heterogeneous and may present strong discontinuities in refractive index, for example at interfaces. Future investigations into Raman assessment of refractive index in human specimens will be needed to confirm the monitoring capacity of the model. For instance, it remains to be seen how similar approaches can be applied to generate refractive index maps overlaid to Brillouin-Raman images of biological specimens. However, the proof of principle presented here shows great potential for future quantitative Brillouin elastography.

**Declaration of Conflicting Interests**

The authors declare no conflicts of interest.

**Funding**

This work was supported by Cancer Research UK/Engineering and Physical Sciences Research Council (NS/A000063/1), Engineering and Physical Sciences Research Council (EP/M028739/1) and COST Action "BioBrillouin" (CA16124).